\documentclass[12pt,epsfig]{article}
\usepackage{amssymb,epsfig}

\newcommand{\half} {\frac{1}{2}}
\newcommand{\quart} {\frac{1}{4}}
\newcommand{\abs} [1] {\vert#1\vert}
\newcommand{\im} {\mathrm{Im}}
\newcommand{\sgn} {\mathrm{sgn}}

\newcommand{\Ahat} {\widehat{A}}

\begin{document}

\begin{titlepage}
\hfill
\vbox{
    \halign{#\hfil         \cr
           } 
      }  
\vspace*{20mm}

\begin{center}
{\Large {\bf Large Charge Four-Dimensional\\ Extremal $N=2$ Black Holes with $R^2$-Terms}\\} \vspace*{15mm}

{\sc Eyal Gruss} \footnote{e-mail: {\tt eyalgruss@gmail.com}} and
{\sc Yaron Oz} \footnote{e-mail: {\tt yaronoz@post.tau.ac.il}}

\vspace*{1cm}
{\it Raymond and Beverly Sackler School of Physics and Astronomy,\\
Tel-Aviv University, Tel-Aviv 69978, Israel.\\}

\end{center}

\vspace*{8mm}

\begin{abstract}

We consider $N=2$ supergravity in four dimensions with small $R^2$ curvature corrections. We construct large charge extremal supersymmetric and non-supersymmetric black hole solutions in all space, and analyze their thermodynamic properties.

\end{abstract}
\vskip 0.8cm

February 2009

\end{titlepage}

\setcounter{footnote}{0}

\section{Introduction}

Charged black holes in four-dimensional $N=2$ supergravity have attracted much attention in recent years. In particular, it is expected that the (indexed) entropy of $N=2$ supersymmetric black holes may be understood from a microscopic construction. From the macroscopic viewpoint, it is of much interest to understand the higher curvature corrections. Higher curvature corrections in regular Einstein gravity black hole solutions and their thermodynamic properties, have been studied as of the mid-1980's \cite{CMP}. There, a solution in all space is obtained by treating the higher curvature term as a small perturbation. For higher curvature corrections of four-dimensional $N=2$ supergravity, the horizon geometry and entropy of supersymmetric black holes have been studied a decade ago \cite{r2entropy, mohauptreview}. The same have been studied recently, for extremal non-supersymmetric black holes \cite{nonsusy1,nonsusy2}. Several works have considered the interpolation of supersymmetric horizon solutions in this theory to asymptotically flat space at infinity \cite{r2stabeq,r2simple,r2interpolation,SJR}. In this work we consider $N=2$ supergravity in four dimensions with small $R^2$ curvature corrections. We construct large charge extremal supersymmetric and non-supersymmetric black hole solutions in all space, and analyze their thermodynamic properties.

Note that the $R^2$-terms considered in this paper are $F$-terms. One generally expects also $D$-term corrections, which are not taken into account here. For supersymmetric black holes, it is conjectured that such terms do not contribute to the entropy \cite{osv}.

The paper is organized as follows: In section $2$ we give a brief review of four-dimensional $N=2$ supergravity with $R^2$-terms. In section $3$ we discuss the framework and method for constructing the large charge black hole solutions. In section $3.1$ we present the solutions and their thermodynamic properties in the case of supersymmetric black holes with $R^2$-terms. In section $3.2$ we do the same for extremal non-supersymmetric black holes with $R^2$-terms.

In the paper we will use $a,b,\ldots=0,1,2,3$ to denote the tangent space
indices, corresponding to the indices $\mu,\nu,\ldots$ of the space-time
coordinates $(t,r,\phi,\theta)$. The exception are $i,j=1,2$ which are gauge $SU(2)$ indices, and $\alpha=1,2$ which is a global $SU(2)$ indices. The sign conventions for the curvature tensors follow \cite{mohauptreview}.

\section{$R^2$-Terms in $N=2$ Supergravity - A Brief Review}

We will consider $N=2$ Poincar\'e supergravity coupled to $N_V$
Abelian $N=2$ vector multiplets. $N=2$ Poincar\'e supergravity is a supersymmetric extension of
Einstein-Maxwell gravity, adding two spin $3/2$ gravitini to the
graviton and (gravi-)photon. $N=2$ Poincar\'e supergravity can
be formulated as a gauge fixed version of $N=2$ conformal
supergravity coupled to an $N=2$ Abelian vector multiplet (see
\cite{mohauptreview} for a comprehensive review).

The on-shell field content of the vector multiplet is a complex
scalar, a doublet of Weyl fermions, and a vector gauge field. We
will consider $N_V+1$ vector multiplets, and will denote by $X^I$,
$I=0\ldots N_V$, the scalars (moduli) in the vector multiplets. The
couplings of the vector multiplets are encoded in a prepotential
$F(X^I)$, which is a homogenous of second degree holomorphic
function.

The $N=2$ conformal supergravity multiplet (Weyl multiplet) is
denoted by $W^{abij}$. It consists
of gauge fields for the local symmetries: translations ($P$),
Lorentz transformations ($M$), dilatations ($D$), special conformal
transformations ($K$), $U(1)$ transformations ($A$), $SU(2)$
transformations ($V$), and supertransformations ($Q,S$). In the
theory without $R^2$-terms, the Weyl multiplet appears in the
Lagrangian through the superconformal covariantizations. In order to
get the $R^2$-terms, one adds explicit couplings to the Weyl
multiplet. This appears in the form of a background chiral multiplet, which is
equal to the square of the Weyl multiplet $W^2$. The lowest
component of the chiral multiplet is a complex scalar denoted
$\Ahat$. The prepotential $F(X^I,\Ahat)$ describes the coupling of
the vector multiplets and the chiral multiplet. One introduces the
notation:
\begin{equation}
F_I\equiv\frac{\partial}{\partial X^I}F(X^I,\Ahat),~~~~~~~~
F_{\Ahat}\equiv\frac{\partial}{\partial \Ahat}F(X^I,\Ahat) \ ,
\end{equation}
and similarly for higher order and mixed derivatives.

The bosonic part of the $N=2$ conformal supergravity Lagrangian is
\begin{eqnarray}
\label{lag} 8\pi
e^{-1}\mathcal{L}&=&-i(\bar{X}^IF_I-X^I\bar{F}_I)(\frac{1}{6}R-D)+{}\nonumber\\*
&&{}+\Big(i\mathcal{D}^a\bar{X}^I\mathcal{D}_aF_I+
\frac{i}{4}F_{IJ}(F_{ab}^{-I}-\quart\bar{X}^IT_{ab}^-)
(F^{ab-J}-\quart\bar{X}^JT^{ab-})+{}\nonumber\\*
&&{}+\frac{i}{8}\bar{F}_I(F_{ab}^{-I}-\quart\bar{X}^IT_{ab}^-)T^{ab-}+
\frac{i}{32}\bar{F}T_{ab}^-T^{ab-}-\frac{i}{8}F_{IJ}Y_{ij}^IY^{ijJ}+{}\nonumber\\*
&&{}-\frac{i}{8}F_{\Ahat\Ahat}
(\widehat{B}_{ij}\widehat{B}^{ij}-2\widehat{F}_{ab}^-\widehat{F}^{ab-})+
\frac{i}{2}\widehat{F}^{ab-}F_{\Ahat
I}(F_{ab}^{-I}-\quart\bar{X}^IT_{ab}^-)+{}\nonumber\\*
&&{}-\frac{i}{4}\widehat{B}_{ij}F_{\Ahat I}Y^{ijI}+
\frac{i}{2}F_{\Ahat}\widehat{C}+\mathrm{h.c.}\Big)+L_{cm} \ .\nonumber\\*
\end{eqnarray}
$e\equiv\sqrt{\abs{\mathrm{det}(g_{\mu\nu})}}$ where $g_{\mu\nu}$
is the curved metric, $R$ is the Ricci scalar, $D$ is an
auxiliary real scalar field of the Weyl multiplet, $D_a$ is
the covariant derivative with respect to all superconformal
transformations, $\mathcal{D}_a$ is the covariant derivative with
respect to $P,M,D,A,V$-transformations, $F_{ab}^{-I}$ is the
anti-selfdual part of the vector field strength, $T_{ab}^-$ is an
anti-selfdual antisymmetric auxiliary field of the Weyl multiplet, and $Y_{ij}^I$ are real $SU(2)$ triplets of auxiliary
scalars of the vector multiplet. The hatted fields are components of the
chiral multiplet $W^2$, with their bosonic parts given by
\begin{eqnarray}
\label{hatted} \theta^0~~~~~~~~\Ahat&=&T_{ab}^-T^{ab-}\nonumber\\*
\theta^2~~~~~~\widehat{B}_{ij}&=&-16R(V)_{(ij)ab}T^{ab-}\nonumber\\*
\widehat{F}^{ab-}&=&-16\mathcal{R}(M)_{cd}^{\phantom{cd}ab}T^{cd-}\nonumber\\*
\theta^4~~~~~~~~
\widehat{C}&=&64\mathcal{R}(M)_{cd}^{\phantom{cd}ab-}\mathcal{R}(M)^{cd-}_{\phantom{cd}ab}+
32R(V)_{ab\phantom{i}j}^{\phantom{ab}i-}R(V)^{abj-}_{\phantom{abj}i}-16T^{ab-}D_aD^cT_{cb}^+
\ .\nonumber\\*
\end{eqnarray}
$T_{ab}^+=\bar{T}_{ab}^-$ is the selfdual counterpart of the
auxiliary field, $R(V)_{ab\phantom{i}j}^{\phantom{ab}i}$ is the
field strength of the $SU(2)$ transformations,
$\mathcal{R}(M)_{ab}^{\phantom{ab}cd}$ is the modified Riemann
curvature and $\mathcal{R}(M)_{ab}^{\phantom{ab}cd-}$ is the
anti-selfdual projection in both pairs of indices. The bosonic part
of $\mathcal{R}(M)_{ab}^{\phantom{ab}cd}$ is given by\footnote{We
have assumed the $K$-gauge fixing which will be defined later
(\ref{gauge}).}
\begin{equation}
\mathcal{R}(M)_{ab}^{\phantom{ab}cd}=R_{ab}^{\phantom{ab}cd}-
4f_{[a}^{\phantom{[a}[c}\delta_{b]}^{d]}+
\frac{1}{32}(T_{ab}^-T^{cd+}+T_{ab}^+T^{cd-}) \ ,
\end{equation}
where $R_{ab}^{\phantom{ab}cd}$ is the Riemann tensor, and
$f_a^{\phantom{a}c}$ is the connection of the special conformal
transformations, determined by the conformal supergravity
conventional constraints, with the bosonic
part\footnotemark[\value{footnote}]:
\begin{equation}
f_a^{\phantom{a}c}=\half
R_a^{\phantom{a}c}-\quart(D+\frac{1}{3}R)\delta_a^c+\frac{1}{2}{}^{\star}R(A)_a^{\phantom{a}c}+
\frac{1}{32}T_{ab}^-T^{cb+} \ ,
\end{equation}
where $R_a^{\phantom{a}c}$ is the Ricci tensor, and
${}^{\star}R(A)^a_{\phantom{a}b}$ is the Hodge dual of the field
strength of the $U(1)$ transformations. Note that the $T^2$-terms
in $\mathcal{R}(M)_{ab}^{\phantom{ab}cd}$ cancel exactly the $T^2$
contribution from $f_a^{\phantom{a}c}$. Finally, $L_{cm}$ denotes additional terms belonging to a compensating multiplet, that add some missing degrees of freedom needed for Poincar\'e supergravity, and take care of the $D$ field appearing linearly in the Lagrangian. We discuss two equivalent possibilities: a nonlinear multiplet and a hypermultiplet.

The bosonic content of the nonlinear multiplet is a real vector field $V_a$, a complex antisymmetric $SU(2)$ triplet of scalars $M_{ij}$, and an $SU(2)$ matrix of scalars $\Phi^i_{\phantom{i}\alpha}$. These are subject to the constraint:
\begin{equation}
\label{nl}
-\frac{1}{3}R-D+\mathcal{D}^aV_a-\half V^aV_a-\quart
M_{ij}\bar{M}^{ij}+D^a\Phi^i_{\phantom{i}\alpha}D_a\Phi^{\alpha}_{\phantom{\alpha}i}=0
\ .
\end{equation}
In the absence of $R^2$-terms this constraint may be imposed by
\begin{equation}
L_{cm}=i(\bar{X}^IF_I-X^I\bar{F}_I)\Big(-\frac{1}{3}R-D+\mathcal{D}^aV_a-\half V^aV_a-
\quart M_{ij}\bar{M}^{ij}+D^a\Phi^i_{\phantom{i}\alpha}D_a\Phi^{\alpha}_{\phantom{\alpha}i}\Big) \ .
\end{equation}
This cancels the $D$ field in the first line of the Lagrangian (\ref{lag}), and leaves $-i(\bar{X}^IF_I-X^I\bar{F}_I)R/2$ as the Einstein-Hilbert term.
In the presence of $R^2$-terms the above $L_{cm}$ does not suffice, and one must also substitute the $D$ field appearing in the hatted fields (\ref{hatted}) in the Lagrangian using (\ref{nl}).

Alternatively, one may use a hypermultiplet as the compensating multiplet \cite{r2stabeq}. The bosonic content of the hypermultiplet are scalars $A_i^{\phantom{i}\Gamma}$ with $\Gamma,\Delta=1\ldots2r$ and $r$ is an integer. We have
\begin{equation}
L_{cm}=-\half\bar{\Omega}_{\Gamma\Delta}\mathcal{D}_aA_i^{\phantom{i}\Gamma}\mathcal{D}^aA^{i\Delta}+\half\chi(\frac{1}{3}R+D) \ ,
\end{equation}
where $\bar{\Omega}_{\Gamma\Delta}$ is a covariantly constant antisymmetric tensor, and the hyper-K\"ahler potential $\chi$ is defined by
\begin{equation}
\chi=\half\bar{\Omega}_{\Gamma\Delta}A_i^{\phantom{i}\Gamma}A^{i\Delta} \ .
\end{equation}

In order to obtain Poincar\'e supergravity one gauge fixes the
bosonic fields (in addition there is a gauge fixing of fermionic
fields):
\begin{eqnarray}
K\mathrm{-gauge}:&&b_a=0\nonumber\\*
A\mathrm{-gauge}:&&X^0=\bar{X}^0>0\nonumber\\*
V\mathrm{-gauge}:&&\Phi^i_{\phantom{i}\alpha}=\delta_{\alpha}^i\qquad\qquad\qquad\quad\textrm{(nonlinear multiplet)}\nonumber\\*
&&A_i^{\phantom{i}\Gamma}=\delta^i_{\Gamma}\sqrt{\half A_j^{\phantom{j}\Delta}A^j_{\phantom{j}\Delta}}\qquad\textrm{(hypermultiplet)} \ ,
\label{gauge}
\end{eqnarray}
where $b_a$ is the connection of the dilatations. Note that we have left out the $D$-gauge, which will be discussed later.

\section{Large Charge $N=2$ Black Hole Solutions\\ with $R^2$-Terms}

We assume a prepotential of the form:
\begin{equation}
\label{prep}
F(X,\Ahat)=\frac{D_{ABC}X^AX^BX^C}{X^0}+\epsilon\frac{D_AX^A}{X^0}\Ahat \ ,
\end{equation}
where $D_{ABC},D_A$ are constants, $D_{ABC}$ is symmetric in all indices, and $A,B,C=1\ldots N_V$. The second term describes $R^2$ couplings in the Lagrangian. It may arise as a $g_s$ correction in the large volume limit\footnote{The large Calabi-Yau volume approximation requires $\im(X^A/X^0)\gg1$.} of type IIA string theory compactified on a Calabi-Yau three-fold, or as an $\alpha'$ correction in heterotic string theory compactified on $K3\times T^2$. We will treat the higher curvature terms in the Lagrangian as a small perturbation, in the spirit of \cite{CMP}. This is valid for the exterior region of black hole solutions in the large charge approximation. The physical dimensionless expansion parameter is one over charge squared: $Q^{-2}$. It is however convenient to express this as an expansion in $\abs{D_A}\ll\abs{D_{ABC}}$ (see also \cite{r2simple,nonsusy1,nonsusy2}). To make this explicit, we have inserted the expansion parameter $\epsilon$ in front of the second term in (\ref{prep}), and at the end we will set $\epsilon=1$.

We look for static spherically symmetric solutions, where the metric takes the form:
\begin{equation}
ds^2=-e^{-2U(r)}f(r)dt^2+e^{2U(r)}\left(f(r)^{-1}dr^2+r^2\sin^2{\theta}d\phi^2+r^2d\theta^2\right) \ .
\label{metric}
\end{equation}
Introduce the rescaled variables:
\begin{eqnarray}
\label{scaling}
Y^I&=&e^UX^I\nonumber\\*
\Upsilon&=&e^{2U}\Ahat\nonumber\\*
e^{-{K}}&=&i\left(\bar{Y}^IF_I(Y,\Upsilon)-Y^I\bar{F}_I(\bar{Y},\bar{\Upsilon})\right) \ .
\end{eqnarray}
The latter is called the K\"ahler potential.\footnote{The scaling $X(z)^I=e^{-{K}/2}X^I$ used in some previous works is not general enough.}

Consider black holes with one electric charge $q_0$ and $p^A$ ($A=1,2,3$) magnetic charges. In our convention $\mathcal{D}\equiv D_{ABC}p^Ap^Bp^C>0$.\footnote{The common convention uses $\mathcal{D}<0$ and a reversed sign for $q_0$.} For $q_0>0$ one can have a supersymmetric solution \cite{genstabeq,r2entropy,SJR}. By reversing the sign of the charge $q_0$, but taking the moduli to depend on the absolute value, one can have an extremal non-supersymmetric solution \cite{sign_reversal,nonsusy1,nonsusy2}. At the $R$-level, i.e. without $R^2$-terms, in addition to the sign changes, the supersymmetric and non-supersymmetric solutions differ also in the form of the auxiliary $T$ field. The thermodynamic properties of the non-supersymmetric solution can be obtained by an analytic continuation. This is no longer true when including higher curvature corrections. We will construct the $R^2$-level solutions for both cases. In the following we will denote $E\equiv D_Ap^A$.

We are interested in black hole solutions of the $R^2$ curvature corrected theory, to first order in $\epsilon$. As a starting point for our ansatz, we may take the $R$-level solution ($\epsilon=0$), with the prepotential $F(\epsilon=0)$ replaced by $F(\epsilon)$. This, however, proves to be insufficient, and we need to introduce a further general linear $\epsilon$-correction to the fields. We look for solutions in the form:
\begin{eqnarray}
\label{r2sol}
e^{2U}&=&e^{-K(\epsilon)}(1+\epsilon\xi_U(r))\nonumber\\*
f&=&1+\epsilon\xi_f(r)\nonumber\\*
Y^A&=&-\frac{i}{2}y^A(1+\epsilon\xi_A(r))\qquad\textrm{(no
summation)}\nonumber\\*
Y^0&=&\half\sqrt{\frac{D_{ABC}y^Ay^By^C-4\epsilon D_Ay^A\Upsilon}{y_0}}(1+\epsilon\xi_0(r))\nonumber\\*
T_{01}^-&=&iT_{23}^-=\left(\frac{3k^3}{r+k^3}+\sgn(q_0)\frac{k_0}{r+k_0}\right)\frac{1}{r}e^{K(\epsilon)/2}(1+\epsilon\xi_T(r)) \ ,
\end{eqnarray}
where $(k_0,k^A)>0$ are constants with either $k^1=k^2=k^3$ or otherwise only $D_{333}\neq0$, and
\begin{eqnarray}
y^A&\equiv&\frac{p^A}{k^A}+\frac{p^A}{r}\qquad\textrm{(no
summation)}\nonumber\\*
y_0&\equiv&\frac{\abs{q_0}}{k_0}+\frac{\abs{q_0}}{r} \ .
\end{eqnarray}
We use
\begin{equation}
\Upsilon=-4e^{2U}(T_{01}^-)^2=-4\left(\frac{3k^3}{r+k^3}+\sgn(q_0)\frac{k_0}{r+k_0}\right)^2\frac{1}{r^2}+O(\epsilon) \ ,
\end{equation}
which is a sufficient approximation since $\Upsilon$ always comes with a factor of $\epsilon$.

The event horizon is located at $r=0$. In order for the perturbative expansion to be valid, we require $\abs{\epsilon\xi(r)}\ll1$ for all $\xi$-functions for $r\geq0$. In addition we set the boundary conditions: $\lim_{r\rightarrow\infty}\xi(r)=0$. This gives an asymptotically flat solution.

Let us introduce the dual field strength:
\begin{equation}
G_{abI}^-=2i\frac{\partial(e^{-1}\mathcal{L})}{\partial F^{ab-I}}=F_{IJ}F_{ab}^{-J}+\quart(\bar{F}_I-F_{IJ}\bar{X}^J)T_{ab}^-+F_{\Ahat I}\widehat{F}_{ab}^- \ ,
\end{equation}
where we have considered only bosonic terms. Due to spherical symmetry we have
\begin{eqnarray}
F_{23}^{-I}&=&-iF_{01}^{-I}\nonumber\\*
G_{23I}^-&=&-iG_{01I}^- \ .
\end{eqnarray}
The field strengths $F_{01}^{-I}$ may be extracted from the following equations:
\begin{equation}
2(\im F_{IJ})F_{01}^{-J}=G_{23I}-\bar{F}_{IJ}F_{23}^J+\half\im\Big(\left(F_I+F_{IJ}\bar{X}^J-64F_{\Ahat I}(2C_{0101}-D)\right)T_{01}^-\Big) \ ,
\end{equation}
where we used spherical symmetry and
\begin{equation}
\mathcal{R}(M)^{mn}_{\phantom{mn}pq}=C^{mn}_{\phantom{mn}pq}+D\delta_{[p}^{[m}\delta_{q]}^{n]}-
2\delta_{[p}^{[m}{}^{\star}R(A)^{n]}_{\phantom{n]}q]} \ ,
\end{equation}
where $C^{ab}_{\phantom{ab}cd}$ is the Weyl tensor. The magnetic parts of the field strengths are obtained from Bianchi identities, which for our static spherically symmetric metric (\ref{metric}) give:
\begin{eqnarray}
F_{23}^I&=&\frac{1}{r^2}e^{-2U}p^I\nonumber\\*
G_{23I}&=&\frac{1}{r^2}e^{-2U}q_I \ .
\end{eqnarray}
For our choice of charges, and the complex-valued form of the prepotential (\ref{prep}) and the ansatz (\ref{r2sol}) we get
\begin{eqnarray}
F_{01}^{-0}&=&\frac{1}{2F_{00}}\Big(iG_{230}-iF_{0A}F_{23}^A+\half\left(F_0+F_{0I}\bar{X}^I-64F_{\Ahat0}(2C_{0101}-D)\right)T_{01}^-\Big)\nonumber\\*
F_{01}^{-A}&=&\frac{i}{2}F_{23}^A \ .
\end{eqnarray}

For the $U(1)$ and $SU(2)$ connections we assume
\begin{eqnarray}
A_a&=&0\nonumber\\*
\mathcal{V}_{a\phantom{i}j}^{\phantom{a}i}&=&0 \
.
\end{eqnarray}
The equation of motion for the $SU(2)$ connection is always satisfied by the vanishing $SU(2)$ connection, for a bosonic background and with either the $V$-gauge for the nonlinear multiplet (\ref{gauge}) or covariantly constant hypermultiplet scalars. This is because the $SU(2)$ connection and its derivatives, appear then in the Lagrangian (\ref{lag}) always in at least a quadratic form.\footnote{More generally, the $SU(2)$ field equations are automatically satisfied since the solution is a singlet under the $SU(2)$ symmetry.} The vanishing of the $SU(2)$ connection implies also $Y_{ij}^I=0$ \cite{yeom}.

When using the nonlinear multiplet, the auxiliary field $D$ may be determined by the constraint on the
nonlinear multiplet (\ref{nl}). We assume
\begin{eqnarray}
V_a&=&0\nonumber\\*
M_{ij}&=&0\nonumber\\*
\Phi^i_{\phantom{i}\alpha}&=&\delta_{\alpha}^i \ ,
\end{eqnarray}
where the later equation is the $V$-gauge.
The equations of motion for $M_{ij}$ and $\Phi^i_{\phantom{i}\alpha}$ are trivially satisfied by the above assumption, where for the latter we assume a vanishing $SU(2)$ connection.
We therefore remain with the constraint:
\begin{equation}
D=-\frac{1}{3}R \ .
\label{DR}
\end{equation}

When using the hypermultiplet, we assume covariantly constant hypermultiplet scalars:
\begin{equation}
\mathcal{D}_aA_i^{\phantom{i}\Gamma}=0 \ .
\end{equation}
The equation of motion for the hyperscalars then gives (\ref{DR}).
Solving the other equations of motion we will find that for all our cases:
\begin{equation}
\chi=-2+O(\epsilon^2) \ .
\end{equation}

For our ansatz to constitute a solution, it must satisfy the
equations of motion for the metric\footnote{For a discussion on the derivation of the metric field equations, see \cite[appendix B]{r2ne}.}, the moduli
$Y^I$, the auxiliary field $T_{01}^-$, the $U(1)$
connection $A_a$, and either $V_a$ for the nonlinear multiplet or the auxiliary field $D$ for the hypermultiplet.
In the cases that we solved, we observed that when using the hypermultiplet, the equation of motion for $D$ has an overall factor of $(k^3-k_0)^2$, after substituting the ansatz. For $k^3=k_0$, one has to take this limit only after solving the equations of motion, in order not to lose a constraint.

From the Einstein-Hilbert term in the Lagrangian (\ref{lag}), one sees that ``Newton's constant'' is given by the unscaled K\"ahler potential:
\begin{equation}
G_N^{-1}=e^{-\mathcal{K}}=i\left(\bar{X}^IF_I(X,\Ahat)-X^I\bar{F}_I(X,\Ahat)\right)=1-\epsilon\xi_U(r) \ .
\end{equation}
Usually one fixes $G_N=1$ as the dilatational $D$-gauge choice. This, however, is too restrictive and does not always allow a solution. Therefore $G_N$ is a function of the radial coordinate, resembling the case of dilaton gravity. The metric in the Einstein frame is given by
\begin{equation}
g^E_{\mu\nu}=G_N^{-1}g_{\mu\nu} \ .
\end{equation}
The ADM mass (in Planck units) for a non-normalized metric is given by
\begin{equation}
g^E_{tt}\big|_{r\rightarrow\infty}=g^E_{tt}(\infty)\left(1-g^E_{rr}(\infty)^{-1/2}\frac{2M}{r}+O(\frac{1}{r^2})\right) \ .
\end{equation}
One may see this by applying the coordinate transformation $t\rightarrow(-g^E_{tt}(\infty))^{-1/2}t,~r\rightarrow g^E_{rr}(\infty)^{-1/2}r$, to get the conventionally normalized line element.

The central charge is the conserved charge associated with the graviphoton:
\begin{eqnarray}
Z&=&\frac{1}{4\pi}\oint_{S^2_\infty}e^{\mathcal{K}/2}(F^{-I}F_I(X,\Ahat)-G^-_IX^I)=\frac{1}{4\pi}\oint_{S^2_\infty}e^{\mathcal{K}/2}(F^IF_I(X,\Ahat)-G_IX^I)=\nonumber\\*
&=&\lim_{r\rightarrow\infty}e^{\mathcal{K}/2}(p^IF_I(X,\Ahat)-q_IX^I)=\lim_{r\rightarrow\infty}e^{K/2}(p^IF_I(Y,\Upsilon)-q_IY^I) \ ,
\end{eqnarray}
where the first equality is valid due to the $N=2$ supersymmetry of flat space at infinity. This also allows to express $Z$ in terms of the auxiliary $T$ field\footnote{The $T$ field differs from the graviphoton $e^{\mathcal{K}}(F^{-I}F_I-G^-_IX^I)$, when $R^2$-terms are included.}:
\begin{equation}
Z=-\frac{i}{16\pi}\oint_{S^2_\infty}e^{-\mathcal{K}/2}T^-=-\frac{1}{4}\lim_{r\rightarrow\infty}e^{-\mathcal{K}/2}r^2e^{2U}T_{01}^- \ .
\end{equation}
$Z$ is determined by the charges and the asymptotic moduli values at infinity, and does not receive higher curvature corrections.
It reads
\begin{equation}
\label{Z}
\abs{Z}=\frac{\sqrt{2}}{4}(\abs{h_0}D_{ABC}h^Ah^Bh^C)^{1/4}\abs{3k^3+\sgn(q_0)k_0}=\quart\abs{3k^3+\sgn(q_0)k_0} \ ,
\end{equation}
where
\begin{equation}
(h_0,h^A)\equiv\left(\frac{q_0}{k_0},\frac{p^A}{k^A}\right)\qquad\textrm{(no summation)} \ ,
\end{equation}
and where in the second equality of (\ref{Z}) we imposed the normalization $g_{tt}(\infty)=-1$:
\begin{equation}
4\abs{h_0}D_{ABC}h^Ah^Bh^C=1 \ .
\end{equation}
The supersymmetry algebra requires $M\geq\abs{Z}$.

Our calculations were done using Maple with GRTensor.

\subsection{Supersymmetric Black Holes}

In the supersymmetric case ($q_0>0$) with $k^1=k^2=k^3$, the solution reads
\begin{eqnarray}
\xi_f(r)&=&\xi_0(r)=\xi_1(r)=\xi_2(r)=\xi_3(r)=0\nonumber\\*
\xi_U(r)&=&\frac{8Er(k^3)^2(k^3-k_0)(9rk^3-rk_0+8k^3k_0)}{\mathcal{D}(r+k^3)^4(r+k_0)^2}\nonumber\\*
\xi_T(r)&=&\frac{4Er(k^3)^2(k^3-k_0)P(r)}{\mathcal{D}(r+k^3)^4(r+k_0)^2(3rk^3+rk_0+4k^3k_0)} \ ,
\end{eqnarray}
where
\begin{equation}
P(r)=144r^3k^3-16r^3k_0-27r^2(k^3)^2+226r^2k^3k_0-7r^2k_0^2-100r(k^3)^2k_0+100rk^3k_0^2-64(k^3)^2k_0^2 \ .
\end{equation}
This can be obtained either from the full second order equations of motion, or from the first order $N=1$ supersymmetry equations \cite{r2stabeq} similarly to \cite{SJR}\footnote{The case discussed in \cite{SJR} with $D_{ABC}=D_{123}$ and constant $Y^1,Y^2$, is not soluble in our approximation scheme.}.
At the horizon ($r=0$) we have:
\begin{equation}
\label{susy_horizon}
\xi_U(0)=\xi_T(0)=0 \ .
\end{equation}
The horizon solution agrees with previous results \cite{r2entropy}.

The entropy is given by the Wald formula for the supersymmetric case with $R^2$-terms \cite{r2entropy}:
\begin{equation}
\label{susy_entropy}
S=\lim_{r\rightarrow0}\left(\frac{A}{4G_N}-4A\cdot\im\left(F_{\Ahat}\abs{T_{01}^-}^2\right)\right) \ ,
\end{equation}
where $A$ is the area of the event horizon. Plugging in the solution gives the expected result:
\begin{equation}
S=2\pi\sqrt{q_0\mathcal{D}}\left(1+\frac{128E}{\mathcal{D}}\right)+O(Q^{-2})\approx2\pi\sqrt{q_0(\mathcal{D}+256E)} \ .
\end{equation}
In the supersymmetric case the ADM mass saturates the BPS bound:
\begin{equation}
M=\abs{Z}=\quart(3k^3+k_0) \ .
\end{equation}
This result is exact and is identical to that of the $R$-level.

Note that in the special case $k^3=k_0$, the $\xi$-functions vanish in all space. This means that the $R^2$-level solution is simply the $R$-level ansatz with $F(\epsilon=0)$ replaced by $F(\epsilon)$. One may ask whether this behavior continues to higher orders in $\epsilon$. Assuming that the prepotential (\ref{prep}) itself does not contain higher orders in $\epsilon$, the solution to order $\epsilon^2$ would be
\begin{eqnarray}
\xi_f(r)&=&\xi_1(r)=\xi_2(r)=\xi_3(r)=\xi_0(r)=0\nonumber\\*
\xi_U(r)&=&\epsilon\frac{65536E^2rk_0^7}{\mathcal{D}^2(r+k_0)^8}\nonumber\\*
\xi_T(r)&=&\epsilon\frac{32768E^2rk_0^6(7r-k_0)}{\mathcal{D}^2(r+k_0)^8} \ .
\end{eqnarray}

\subsection{Non-supersymmetric Black Holes}

For simplicity, we consider the cases with $D_{ABC}=D_{333}$ or $D_{ABC}=D_{123}$, and $D_A=D_3$ and $k^1=k^2=k^3=k_0$. At the $R$-level, the last condition gives the non-supersymmetric version of the double-extremal black hole. We were able to find similar solutions for different combinations of $D_{ABC}$'s and $D_A$'s, providing that for each term such as $D_3$ there is a at least one corresponding term $D_{AB3}$ (this was not required in the supersymmetric case). We then give the generalization of the $D_{ABC}=D_{333}$ case to arbitrary $k's$.

In the non-supersymmetric case ($q_0<0$) with $k^1=k^2=k^3=k_0$, the solution reads
\begin{eqnarray}
\xi_U(r)&=&\frac{64Ek_0^4}{\mathcal{D}(r+k_0)^4}\nonumber\\*
\xi_f(r)&=&\frac{8Ek_0(4r^3+25r^2k_0+60rk_0^2+30k_0^3)}{15\mathcal{D}(r+k_0)^4}\nonumber\\*
\xi_0(r)&=&\frac{4E\left(-120r(r+k_0)^3\ln\frac{r+k_0}{r}+118r^3k_0+297r^2k_0^2+224rk_0^3+240k_0^4\right)}{15\mathcal{D}(r+k_0)^4}\nonumber\\*
\xi_T(r)&=&\frac{4E\left(120r(r+k_0)^3\ln\frac{r+k_0}{r}-118r^3k_0-297r^2k_0^2+736rk_0^3-120k_0^4\right)}{15\mathcal{D}(r+k_0)^4} \ .
\end{eqnarray}
For $D_{ABC}=D_{333}$ and $D_A=D_3$:
\begin{equation}
\xi_3(r)=\frac{4E\left(-120r(r+k_0)^3\ln\frac{r+k_0}{r}+108r^3k_0+267r^2k_0^2+194rk_0^3+270k_0^4\right)}{45\mathcal{D}(r+k_0)^4} \ ,
\end{equation}
and $\xi_1(r),\xi_2(r)$ are irrelevant.
For $D_{ABC}=D_{123}$ and $D_A=D_3$:
\begin{eqnarray}
\xi_3(r)&=&\frac{4E\left(-840r(r+k_0)^3\ln\frac{r+k_0}{r}+856r^3k_0+2169r^2k_0^2+1658rk_0^3+150k_0^4\right)}{15\mathcal{D}(r+k_0)^4}\nonumber\\*
\xi_1(r)&=&\xi_2(r)=\frac{4E\left(360r(r+k_0)^3\ln\frac{r+k_0}{r}-374r^3k_0-951r^2k_0^2-732rk_0^3+60k_0^4\right)}{15\mathcal{D}(r+k_0)^4} \ .
\end{eqnarray}
At the horizon ($r=0$) we have,
\begin{eqnarray}
\xi_U(0)&=&\frac{64E}{\mathcal{D}}\nonumber\\*
\xi_f(0)&=&\frac{16E}{\mathcal{D}}\nonumber\\*
\xi_0(0)&=&\frac{64E}{\mathcal{D}}\nonumber\\*
\xi_T(0)&=&-\frac{32E}{\mathcal{D}} \ .
\end{eqnarray}
For $D_{ABC}=D_{333}$ and $D_A=D_3$:
\begin{equation}
\xi_3(0)=\frac{24E}{\mathcal{D}} \ .
\end{equation}
For $D_{ABC}=D_{123}$ and $D_A=D_3$:
\begin{eqnarray}
\xi_3(0)&=&\frac{40E}{\mathcal{D}}\nonumber\\*
\xi_1(0)&=&\xi_2(0)=\frac{16E}{\mathcal{D}} \ .
\end{eqnarray}
The latter case agrees with \cite{nonsusy2}, where the horizon solutions were obtained by extremizing the entropy function.\footnote{Note that \cite{nonsusy2} uses a different scaling than (\ref{scaling}).} Note that the radial derivatives of our solutions for $\xi_A(r),\xi_0(r),\xi_T(r)$ diverge at $r=0$. The curvature, however, is regular.

The entropy formula that we use in the supersymmetric case (\ref{susy_entropy}), is no longer valid in the non-supersymmetric case.
In the latter case, one can use either
Sen's entropy function method \cite{nonsusy1,nonsusy2}, or the Wald entropy formula for the non-extremal case with $R^2$-terms \cite{r2ne}:
\begin{equation}
S=\lim_{r\rightarrow0}\left(\frac{A}{4G_N}-4A\cdot\im\Big(F_{\Ahat}(\abs{T_{01}^-}^2+16C_{0101}+16D)\Big)\right) \ .
\end{equation}
In fact, in our solutions $C_{0101},D$ do not contribute to the entropy to first order in $\epsilon$, and thus the formula does reduce to the supersymmetric one. Note also that $G_N(0)\neq1$.
Plugging in the solution gives:
\begin{equation}
S=2\pi\sqrt{-q_0\mathcal{D}}\left(1+\frac{40E}{\mathcal{D}}\right)+O(Q^{-2}) \ ,
\end{equation}
as in \cite{nonsusy1,nonsusy2}. As discussed therein, this differs from the statistical microscopic entropy \cite{micro_nonsusy}, due to higher curvature $D$-terms, which are not taken into account here. Using the entropy function method with our first order solutions, one can get the entropy to second order:
\begin{equation}
S=2\pi\sqrt{-q_0\mathcal{D}}\left(1+\frac{40E}{\mathcal{D}}-\frac{8576E^2}{\mathcal{D}^2}\right)+O(Q^{-4}) \ .
\end{equation}
The central charge reads
\begin{equation}
\abs{Z}=\frac{\sqrt{2}}{2}(-q_0\mathcal{D})^{1/4}=\half k_0 \ ,
\end{equation}
and the ADM mass takes the form:
\begin{equation}
M=\sqrt{2}(-q_0\mathcal{D})^{1/4}\left(1-\frac{12E}{5\mathcal{D}}\right)+O(Q^{-3})=k_0\left(1-\frac{12E}{5\mathcal{D}}\right)+O(Q^{-3}) \ .
\end{equation}
Here the mass gets an $R^2$ curvature correction. This correction comes from the functions $\xi_f(r),\xi_0(r),\xi_A(r)$, which vanished in the supersymmetric case.
The higher curvature correction changes the mass in an opposite direction to the entropy. For the mass to decrease as conjectured in \cite{mass_charge}, $\mathcal{D}$ and $E$ must have the same sign. This would also mean that the entropy increases.

In the non-supersymmetric case ($q_0<0$) with $D_{ABC}=D_{333}$ and $D_A=D_3$ and arbitrary $k$'s, the solution is given in the appendix. The horizon limit, and the entropy are the same as in the $k^3=k_0$ case. The central charge reads
\begin{equation}
\abs{Z}=\quart\abs{3k^3-k_0} \ ,
\end{equation}
and the ADM mass takes the form:
\begin{equation}
M=\quart(3k^3+k_0)+\frac{4Ek^3k_0\left(12(k^3)^3k_0\ln\frac{k^3}{k_0}-(k^3-k_0)(3(k^3)^3+13(k^3)^2k_0-5k^3k_0^2+k_0^3)\right)}{\mathcal{D}(k^3-k_0)^5}+O(Q^{-3}) \ .
\end{equation}
The mass correction term has the same behavior as in the $k^3=k_0$ case.

\section*{Acknowledgements}
This work is supported in part by the Israeli Science Foundation center of excellence,
by the Deutsch-Israelische Projektkooperation (DIP), by the US-Israel Binational Science Foundation (BSF), and by the German-Israeli Foundation (GIF).

\appendix
\section{The Non-supersymmetric Solution with Arbitrary $k$'s}

In the non-supersymmetric case ($q_0<0$) with $D_{ABC}=D_{333}$ and $D_A=D_3$ and arbitrary $k$'s, the solution reads
\begin{eqnarray}
\xi_U(r)&=&\frac{8D_3(k^3)^2(9r^2(k^3)^2-2r^2k^3k_0+r^2k_0^2+16r(k^3)^2k_0+8(k^3)^2k_0^2)}{D_{333}(p^3)^2(r+k^3)^4(r+k_0)^2}\nonumber\\*
\xi_f(r)&=&\frac{16D_3(k^3)^2k_0\left(2k^3(r+k^3)^2P_1(r)\ln\frac{k^3(r+k_0)}{k_0(r+k^3)}+r(k^3-k_0)P_2(r)\right)}{D_{333}(p^3)^2r^2(r+k^3)^2(k^3-k_0)^5}\nonumber\\*
\xi_3(r)&=&\frac{8D_3(-2(r+k^3)^2(r+k_0)L_3(r)+r(k^3)^2k_0(k^3-k_0)P_5(r))}{3D_{333}(p^3)^2r^2k_0(r+k^3)^3(r+k_0)(k^3-k_0)^6}\nonumber\\*
\xi_0(r)&=&\frac{4D_3(2(r+k^3)^2L_0(r)+rk^3k_0(k^3-k_0)P_8(r))}{D_{333}(p^3)^2r^2k^3k_0(r+k^3)^3(r+k_0)(k^3-k_0)^6}\nonumber\\*
\xi_T(r)&=&\frac{4D_3(2(r+k^3)^3L_T(r)+rk^3k_0(k^3-k_0)P_{12}(r))}{D_{333}(p^3)^2r^2k^3k_0(r+k^3)^4(r+k_0)^2(k^3-k_0)^6(3rk^3-rk_0+2k^3k_0)} \ ,
\end{eqnarray}
and $\xi_1(r),\xi_2(r)$ are irrelevant, and where
\begin{eqnarray}
L_3(r)&=&(k^3)^3k_0^2P_3(r)\ln\frac{k^3(r+k_0)}{k_0(r+k^3)}+r^3(k^3)^3P_4(r)\ln\frac{r+k^3}{r+k_0}+r^3(k^3+k_0)(k^3-k_0)^6\ln\frac{r+k^3}{r}\nonumber\\
L_0(r)&=&2(k^3)^4k_0^2P_6(r)\ln\frac{k^3(r+k_0)}{k_0(r+k^3)}+(k^3)^4r^3P_7(r)\ln\frac{r+k^3}{r+k_0}+{}\nonumber\\*
&&{}-r^3(k^3+k_0)(k^3-k_0)^6(rk^3+rk_0+2k^3k_0)\ln\frac{r+k^3}{r}\nonumber\\
L_T(r)&=&2(k^3)^4k_0^2(r+k_0)P_9(r)\ln\frac{k^3(r+k_0)}{k_0(r+k^3)}+2r^3(k^3)^4(r+k_0)P_{10}(r)\ln\frac{r+k^3}{r+k_0}+{}\nonumber\\*
&&{}+2r^3(r+k_0)(k^3+k_0)(k^3-k_0)^6P_{11}(r)\ln\frac{r+k^3}{r}\nonumber\\
P_1(r)&=&2r(k^3)^2+8rk^3k_0+2rk_0^2+3(k^3)^2k_0+9k^3k_0^2\nonumber\\
P_2(r)&=&-12r^2(k^3)^2-12r^2k^3k_0-20r(k^3)^3-25r(k^3)^2k_0-4rk^3k_0^2+rk_0^3-6(k^3)^4-18(k^3)^3k_0\nonumber\\
P_3(r)&=&-6r^2(k^3)^3-6r^2(k^3)^2k_0+66r^2k^3k_0^2+6r^2k_0^3+3r(k^3)^3k_0+30r(k^3)^2k_0^2+27rk^3k_0^3+{}\nonumber\\
&&{}+(k^3)^4k_0+10(k^3)^3k_0^2+9(k^3)^2k_0^3\nonumber\\
P_4(r)&=&-(k^3)^4+5(k^3)^3k_0+9(k^3)^2k_0^2-21k^3k_0^3-12k_0^4\nonumber\\
P_5(r)&=&2r^4(k^3)^4-8r^4(k^3)^3k_0+32r^4(k^3)^2k_0^2+60r^4k^3k_0^3-6r^4k_0^4+4r^3(k^3)^5-21r^3(k^3)^4k_0+{}\nonumber\\*
&&{}+103r^3(k^3)^3k_0^2+143r^3(k^3)^2k_0^3+57r^3k^3k_0^4-6r^3k_0^5+2r^2(k^3)^6-12r^2(k^3)^5k_0+{}\nonumber\\*
&&{}+88r^2(k^3)^4k_0^2+162r^2(k^3)^3k_0^3+126r^2(k^3)^2k_0^4-6r^2k^3k_0^5+10r(k^3)^6k_0-26r(k^3)^5k_0^2+{}\nonumber\\*
&&{}+189r(k^3)^4k_0^3-9r(k^3)^3k_0^4+45r(k^3)^2k_0^5-9rk^3k_0^6+2(k^3)^6k_0^2+20(k^3)^5k_0^3+18(k^3)^4k_0^4\nonumber\\
P_6(r)&=&2r^3(k^3)^3-12r^3(k^3)^2k_0-48r^3k^3k_0^2-2r^3k_0^3-r^2(k^3)^4-15r^2(k^3)^3k_0-42r^2(k^3)^2k_0^2+{}\nonumber\\*
&&{}-59r^2k^3k_0^3-3r^2k_0^4-2r(k^3)^4k_0-22r(k^3)^3k_0^2-38r(k^3)^2k_0^3-18rk^3k_0^4-(k^3)^4k_0^2+{}\nonumber\\*
&&{}-10(k^3)^3k_0^3-9(k^3)^2k_0^4\nonumber\\
P_7(r)&=&r(k^3)^4-4r(k^3)^3k_0-2r(k^3)^2k_0^2+36rk^3k_0^3+9rk_0^4+2(k^3)^4k_0+2(k^3)^3k_0^2+6(k^3)^2k_0^3+{}\nonumber\\*
&&{}+18k^3k_0^4+12k_0^5\nonumber\\
P_8(r)&=&2r^4(k^3)^6-12r^4(k^3)^5k_0+106r^4(k^3)^4k_0^2+76r^4(k^3)^3k_0^3-18r^4(k^3)^2k_0^4+8r^4k^3k_0^5-2r^4k_0^6+{}\nonumber\\*
&&{}+4r^3(k^3)^7-25r^3(k^3)^6k_0+305r^3(k^3)^5k_0^2+228r^3(k^3)^4k_0^3+40r^3(k^3)^3k_0^4+13r^3(k^3)^2k_0^5+{}\nonumber\\*
&&{}-5r^3k^3k_0^6+2r^2(k^3)^8-6r^2(k^3)^7k_0+264r^2(k^3)^6k_0^2+316r^2(k^3)^5k_0^3+138r^2(k^3)^4k_0^4+{}\nonumber\\*
&&{}+10r^2(k^3)^3k_0^5-4r^2(k^3)^2k_0^6+22r(k^3)^8k_0-6r(k^3)^7k_0^2+354r(k^3)^6k_0^3-34r(k^3)^5k_0^4+{}\nonumber\\*
&&{}+80r(k^3)^4k_0^5-16r(k^3)^3k_0^6+4(k^3)^8k_0^2+40(k^3)^7k_0^3+36(k^3)^6k_0^4\nonumber\\
P_9(r)&=&-6r^4(k^3)^4-4r^4(k^3)^3k_0+96r^4(k^3)^2k_0^2+36r^4k^3k_0^3-2r^4k_0^4-5r^3(k^3)^4k_0+81r^3(k^3)^3k_0^2+{}\nonumber\\*
&&{}+249r^3(k^3)^2k_0^3+35r^3k^3k_0^4+3r^2(k^3)^5k_0+46r^2(k^3)^4k_0^2+154r^2(k^3)^3k_0^3+182r^2(k^3)^2k_0^4+{}\nonumber\\*
&&{}+15r^2k^3k_0^5+5r(k^3)^5k_0^2+55r(k^3)^4k_0^3+95r(k^3)^3k_0^4+45r(k^3)^2k_0^5+2(k^3)^5k_0^3+{}\nonumber\\*
&&{}+20(k^3)^4k_0^4+18(k^3)^3k_0^5\nonumber\\
P_{10}(r)&=&-r^2(k^3)^5+5r^2(k^3)^4k_0+8r^2(k^3)^3k_0^2-28r^2(k^3)^2k_0^3-27r^2k^3k_0^4+3r^2k_0^5-2r(k^3)^5k_0+{}\nonumber\\*
&&{}+8r(k^3)^4k_0^2+4r(k^3)^3k_0^3-72r(k^3)^2k_0^4-18rk^3k_0^5-2(k^3)^5k_0^2-2(k^3)^4k_0^3-6(k^3)^3k_0^4+{}\nonumber\\*
&&{}-18(k^3)^2k_0^5-12k^3k_0^6\nonumber\\*
P_{11}(r)&=&r^2(k^3)^2+r^2k_0^2+2r(k^3)^2k_0+2rk^3k_0^2+2(k^3)^2k_0^2\nonumber\\
P_{12}(r)&=&-4r^7(k^3)^7+16r^7(k^3)^6k_0-56r^7(k^3)^5k_0^2-268r^7(k^3)^4k_0^3-20r^7(k^3)^3k_0^4+24r^7(k^3)^2k_0^5+{}\nonumber\\*
&&{}-16r^7k^3k_0^6+4r^7k_0^7-12r^6(k^3)^8+50r^6(k^3)^7k_0-214r^6(k^3)^6k_0^2-1226r^6(k^3)^5k_0^3+{}\nonumber\\*
&&{}-724r^6(k^3)^4k_0^4+86r^6(k^3)^3k_0^5-46r^6(k^3)^2k_0^6+2r^6k^3k_0^7+4r^6k_0^8+132r^5(k^3)^9+{}\nonumber\\*
&&{}-678r^5(k^3)^8k_0+1166r^5(k^3)^7k_0^2-3778r^5(k^3)^6k_0^3-2062r^5(k^3)^5k_0^4-450r^5(k^3)^4k_0^5+{}\nonumber\\*
&&{}-86r^5(k^3)^3k_0^6-22r^5(k^3)^2k_0^7+18r^5k^3k_0^8-31r^4(k^3)^{10}+586r^4(k^3)^9k_0-2706r^4(k^3)^8k_0^2+{}\nonumber\\*
&&{}+2614r^4(k^3)^7k_0^3-9328r^4(k^3)^6k_0^4+822r^4(k^3)^5k_0^5-726r^4(k^3)^4k_0^6-70r^4(k^3)^3k_0^7+{}\nonumber\\*
&&{}+39r^4(k^3)^2k_0^8-70r^3(k^3)^{10}k_0+678r^3(k^3)^9k_0^2-3926r^3(k^3)^8k_0^3+1350r^3(k^3)^7k_0^4+{}\nonumber\\*
&&{}-7394r^3(k^3)^6k_0^5+1922r^3(k^3)^5k_0^6-610r^3(k^3)^4k_0^7+50r^3(k^3)^3k_0^8-80r^2(k^3)^{10}k_0^2+{}\nonumber\\*
&&{}+176r^2(k^3)^9k_0^3-2842r^2(k^3)^8k_0^4+152r^2(k^3)^7k_0^5-2324r^2(k^3)^6k_0^6+728r^2(k^3)^5k_0^7+{}\nonumber\\*
&&{}-130r^2(k^3)^4k_0^8-40r(k^3)^{10}k_0^3-200r(k^3)^9k_0^4-776r(k^3)^8k_0^5-200r(k^3)^7k_0^6+{}\nonumber\\*
&&{}-80r(k^3)^6k_0^7+16r(k^3)^5k_0^8-8(k^3)^{10}k_0^4-80(k^3)^9k_0^5-72(k^3)^8k_0^6 \ .
\end{eqnarray}

\end{document}